\documentclass[amsmath,amssymb,aps,prl,twocolumn,showpacs,superscriptaddress]{revtex4-1}
\usepackage{graphicx}
\usepackage{dcolumn}
\usepackage{bm}
\usepackage{xr}
\makeatletter
\usepackage{hyperref}
\usepackage{xcolor}
\usepackage[stable]{footmisc}
\usepackage{lineno}

\hypersetup{
  colorlinks   = true,%
  urlcolor     = black,
  linkcolor    = blue,
  citecolor   = blue
}

\begin{document}
%\linenumbers
\title{How to make and trap a pseudo-vesicle with a micropipette} %Article title goes here instead of the text "This is the title"

\author{Pierre Tapie}
\affiliation{Sorbonne Universit\'e, CNRS, Institut de Biologie Paris-Seine (IBPS), Laboratoire Jean Perrin (LJP), 4 place Jussieu, F-75005 Paris, France.}

\author{Alexis M. Prevost}
\affiliation{Sorbonne Universit\'e, CNRS, Institut de Biologie Paris-Seine (IBPS), Laboratoire Jean Perrin (LJP), 4 place Jussieu, F-75005 Paris, France.}

\author{Lorraine Montel}
\affiliation{Sorbonne Universit\'e, CNRS, Institut de Biologie Paris-Seine (IBPS), Laboratoire Jean Perrin (LJP), 4 place Jussieu, F-75005 Paris, France.}

\author{L\'ea-Laetitia Pontani}
\email[]{ E-mail: lea-laetitia.pontani@sorbonne-universite.fr}
\affiliation{Sorbonne Universit\'e, CNRS, Institut de Biologie Paris-Seine (IBPS), Laboratoire Jean Perrin (LJP), 4 place Jussieu, F-75005 Paris, France.}

\author{Elie Wandersman}
\email[]{ E-mail: elie.wandersman@sorbonne-universite.fr}
\affiliation{Sorbonne Universit\'e, CNRS, Institut de Biologie Paris-Seine (IBPS), Laboratoire Jean Perrin (LJP), 4 place Jussieu, F-75005 Paris, France.}
 
\begin{abstract}
We present a simple method to produce giant lipid pseudo-vesicles (vesicles with an oily cap on the top), trapped in an agarose gel. The method can be implemented using only a regular micropipette and relies on the formation of a water/oil/water double droplet in liquid agarose. We characterize the produced vesicle with fluorescence imaging and establish the presence and integrity of the lipid bilayer by the successful insertion of $\alpha$-Hemolysin transmembrane proteins. Finally, we show that the vesicle can be easily mechanically deformed, non-intrusively, by indenting the surface of the gel.
\end{abstract}.

\maketitle
\section{Introduction}
Vesicles have been widely used to mimic cellular compartmentalization and reproduce \emph{in vitro} specific biological functions with bottom-up approaches \cite{Wang2021,jeong2020toward}. In many recent studies, the focus was put on the encapsulation of complex biological reactions inside the vesicles, in order to express proteins~\cite{noireaux2004vesicle} or genes~\cite{niederholtmeyer2018communication}, or to reconstitute and study the protein filaments of the cytoskeleton, such as the actin cortex \cite{pontani2009reconstitution} or microtubule asters \cite{Gavriljuk2021}. Many other works aim to mimic the cell membrane properties, as for instance membrane fusion \cite{chan2008lipid,chan2009effects, heuvingh2004hemifusion},  or cell communication, via the insertion of transmembrane proteins in liposomes. In particular, mechanosensitive channels \cite{Garamella2019,Hindley2019} have been inserted in the membrane of liposomes, allowing in turn to measure their conductance under mechanical stress. \\%In this case, the produced mechanosensitive vesicles need to be mechanically excited in a controlled manner. 
The production of vesicles usually relies on either hydration methods or inverted emulsion templates. Hydration methods rely on the swelling of dried lipid films in an aqueous buffer~\cite{bangham1964negative,stein2017production}. They are relatively simple to operate but yield polydisperse vesicle sizes and a relatively non homogeneous encapsulation efficiency. The production rate of unilamellar vesicles can be improved by using electric fields (electroformation methods \cite{angelova1986liposome}) but  fragile proteins can be damaged by the applied fields and the technique is also limited to buffers with low ionic concentrations. Inverted emulsion templates, on the other hand, are based on the forced passage of a water-in-oil emulsion droplets through a water/oil interface~\cite{pontani2009reconstitution,pautot2003production}. This method can be developed in microfluidic chips~\cite{li2018microfluidic}, yielding monodisperse vesicle sizes, that are limited by the microfluidic channel dimensions. \\ 
In both methods (hydration or emulsions), the resulting vesicles are dispersed in the outer medium, which requires additional steps to handle or transfer them in different environments (micropipettes \cite{Evans1987}, optical trapping \cite{Kulin2003}, etc..). These extra steps, relying on specific technical skills, make it difficult to replicate many experiments and collect large statistics on the systems properties. In particular, in order to study mechanotransduction processes, it is required to mechanically stimulate the vesicles. In practice, this has typically been achieved with local membrane deformations (pipette suction \cite{Garten2017}, fluid flows \cite{robinson2019microfluidic,deschamps2009dynamics}, AFM \cite{Schafer2015, Lin2019}). However, these methods do not reproduce faithfully the nature of mechanical perturbations in tissues. In addition, they overlook the mechanical coupling between cells and their biological visco-elastic environment.\\
We propose here a new technique, that provides a versatile platform for the straightforward production of biomimetic pseudo-vesicles (a vesicle with an oily cap on top), but also allows for their trapping and non-intrusive mechanical excitation. 

\section{Pseudo-vesicle formation}
The vesicles are produced from a water in lipid-containing oil droplet configuration, that is formed in a liquid, warm, agarose solution (details on the chemicals can be found in the Methods Section).
We first draw a small volume (about 500 nL) of the internal aqueous phase (Figure 1~a), using a 2.5 $\mu$L micropipette (Eppendorf). Second, we move the pipette into the oil/lipids container and suck about 50 nL of it (Figure 1~b), by turning the adjustment wheel of the micropipette while the tip is immersed in the oil. The pipette tip thus contains an oil/water sandwich. Last, we move the  pipette into the warm agarose solution (temperature $T\approx 38 ^{\circ}C$, see details in the Methods Section) where we expel the oil/water sandwich by turning backward the adjustment wheel of the micropipette. During this expulsion phase, the oil phase first grows into a droplet in the liquid agarose (Figure 1~c), followed by the aqueous phase that grows inside of it (Figure 1~d,e). Finally, we move up the pipette rapidly in the air, which detaches the double droplet from the tip due to viscous friction from the agarose solution (Figure 1~f). \\
Immediately after its formation, the inner water droplet is thus surrounded by an oil layer containing phospholipids. These lipids therefore redistribute at both oil water interfaces, with their hydrophilic heads turned towards the inner water phase on one side and towards the agar solution on the other side. While the agarose solution cools down, the double droplet slowly sediments in the agarose solution and eventually gets trapped in the formed gel. During this cooling process, the oil layer of the double droplet creams under gravity to form an oil cap, while the lipids stabilizing the former oil/water interfaces zip up a lipid bilayer on the lower part of the droplet (Figure 1~g). The double droplet becomes a pseudo-vesicle. This simple micro-pipette method yields pseudo-vesicles with a diameter of 601 $\pm 58 \mu$m (N=98). Smaller sizes can be obtained using smaller tips and a micro-injector (see Methods Section). The rate of success is overall about 30\%. Once trapped in the gel, the pseudo-vesicle is stable for several hours and can be kept overnight if the gel-containing cuvette is sealed to prevent evaporation.

\begin{figure*}[!t]
\centering
\includegraphics[width=0.9\textwidth]{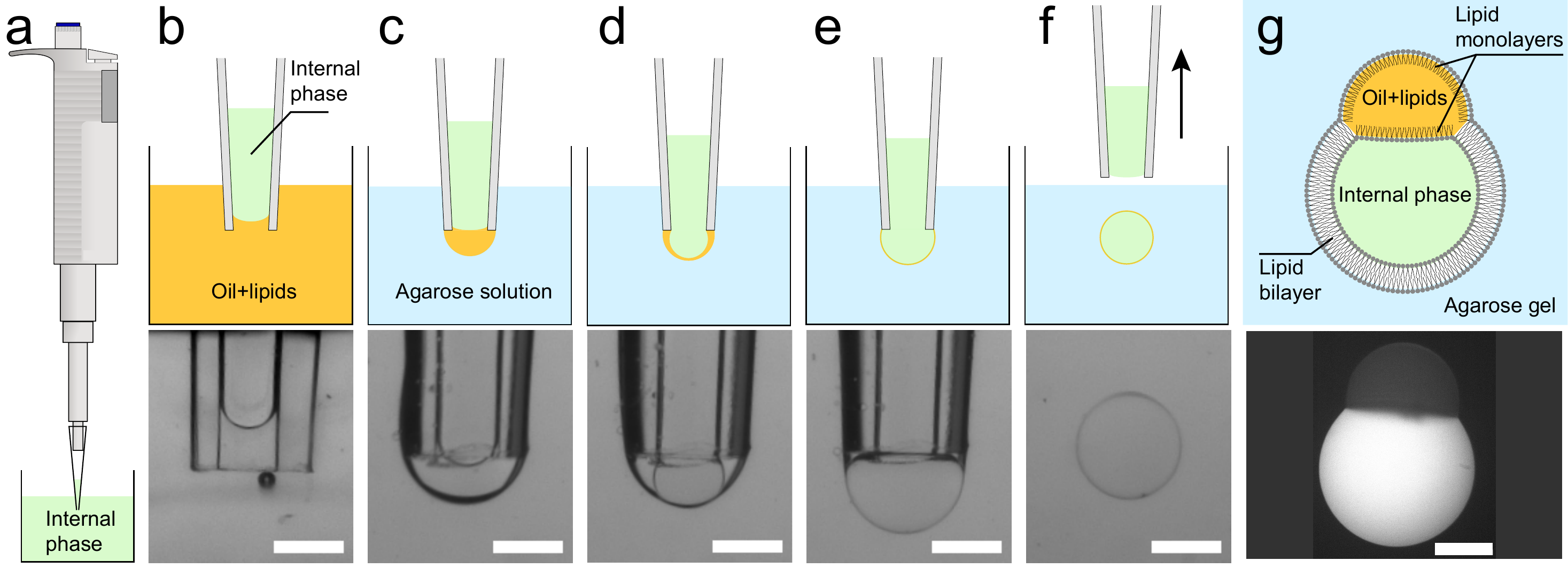}
\caption{a-f) (upper row) Sketch of the double emulsion production method. Green/orange/blue colors stand for internal/oil/external phases, respectively. (bottom row) Bright field images. Scale bars = 500 $\mu$m. g) Top pannel : sketch of the pseudo-vesicle trapped in the agar gel. Bottom pannel: Fluorescence macroscope image of a pseudo-vesicle loaded with carboxyfluorescein trapped in an agar gel. Scale bar = 200 $\mu$m.}
\label{fig:fig1}
\end{figure*}

\section{Membrane characterisation and functionalization}
\subsection{Fluorescence imaging}
\par To characterize the pseudo-vesicle and probe the existence of a lipid bilayer, we first used fluorescent markers dispersed in both the internal aqueous phase (carboxyfluorescein, emission wavelength $\lambda_c = 525$ nm, green, see Methods Section) and the oil phase (Nile red, emission wavelength $\lambda_n = 636$ nm, red). The pseudo-vesicle is produced in the agar gel as described above, and further imaged using a confocal microscope, with a 4x objective. The fluorescence images from both the green and red channels are recorded (see Figure ~2a), showing that the fluorescent water phase is efficiently encapsulated in the pseudo-vesicle. Secondly, we added green fluorescent lipids (3\% wt) to the lipid mixture present in the oil phase labelled with Nile Red. A composite image obtained by confocal microscopy is shown on Figure ~2b. On Figure ~2c, we plot for each channel the normalized radial intensity profiles ($(I(r)-I(0))/(I_{\mathrm{cap}}+I(0))$, where $I_{\mathrm{cap}}$ is the intensity in the oil cap in a given channel. These profiles reveal a significant increase of the fluorescent signal arising from the lipids at the boundary between the inner aqueous phase and the outer agar gel, whereas the fluorescence from the oil phase doesn't significantly increase. Altogether, our results confirm that the inner phase is efficiently encapsulated in a lipid bilayer devoid of significant amounts of oil.

\subsection{Nanopores insertions}
\par To further probe the presence of a lipid bilayer and its functionality, we inserted $\alpha$-Hemolysin ($\alpha$HL) transmembrane proteins in the lipid bilayer. $\alpha$HL is an heptameric nanopore \cite{gouaux1994subunit} through which carboxyfluorescein molecules can diffuse~\cite{valet2019diffusion}. Practically, we found that the direct dissolution of  $\alpha$HL monomers in the internal aqueous phase was decreasing the pseudo-vesicle stability, due to a nanopore induced modification of the oil/water surface tension (See Supplementary Information). We therefore used Small Unilamellar Vesicles (SUV) containing $\alpha$HL nanopores, primarily prepared (see Methods Section) and dispersed in the fluorescent aqueous internal phase loaded with carobxyfluorescein. The pseudo-vesicle is then formed as described above, allowing for the SUVs contained in the inner phase to fuse with the bilayer, thus inserting $\alpha$HL pores into it (Figure ~2d). Under an epifluorescence macroscope, we imaged the carboxyfluorescein leakage across the pseudo-vesicle by acquiring one image every 4 minutes, for two hours. We performed control experiments using on the one hand SUVs devoid of $\alpha$HL nanopores, and on the other hand pseudo-vesicles prepared without SUVs nor $\alpha$HL. Using image analysis, we measured the average intensity inside (\textit{resp.} outside) the pseudo-vesicle $I_{in}$ (\textit{resp.} $I_{out}$) from which we determine the fluorescence intensity contrast $\Gamma = (I_{in}-I_{out})/(I_{in}+I_{out})$. The time variation of the normalized contrast $\Gamma(t) / \Gamma(t=0)$ is presented on Figure ~2e. Clearly, the normalized contrast decreases over time for pseudo-vesicles containing $\alpha$HL whereas it remains constant for $\alpha$HL-free pseudo-vesicles. This establishes that the lipid bilayer can be functionalized with a transmembrane protein. 

 \begin{figure}[!t]
\centering
\includegraphics[width=0.4\textwidth]{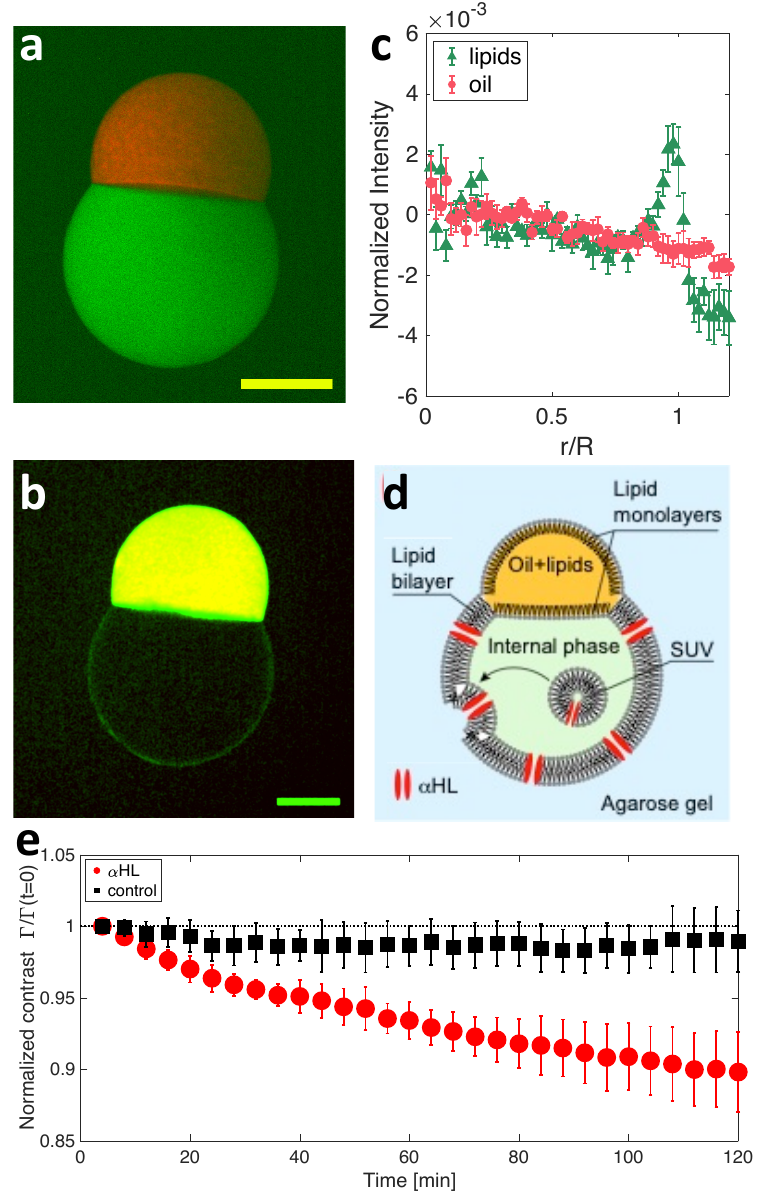}
\caption{a) Composite confocal microscopy image of a pseudo-vesicle trapped in an agar gel. The internal phase contains carboxyfluorescein (green), while the oil phase contains Nile Red (red). b) Composite image using fluorescent lipids (green) and an oil containing Nile Red (red). Yellow color = red + green channels. For clarity, the image has been smoothed with a 2 pixel radius Gaussian filter. Scale bars = 200 $\mu$m. c) Normalized radial fluorescence intensity profile, averaged over N=8 pseudo-vesicles labelled as in b). The green triangles (\textit{resp.} red disks) show the fluorescent lipids channel (\textit{resp.} fluorescent oil channel). Error bars are SE of data. $R=316\pm24$ $\mu$m is the average pseudo-vesicle size. d) Sketch of $\alpha$HL insertion in the pseudo-vesicle membrane, using $\alpha$HL loaded SUV. e) Time evolution of the normalized contrast $\Gamma(t)/\Gamma(t=0)$, for $\alpha$HL loaded pseudo-vesicles (red disks, N=5) and control experiments without $\alpha$HL (black squares, N=5). Error bars are SD of data.}
\label{fig:fig2}
\end{figure}

\section{Mechanical excitation}
 \par Since the pseudo-vesicle is trapped within the bulk of the agarose gel, it can be deformed by indenting the surface of the gel (Figure ~3a).  We compress the surface of the gel with a square piston (surface $S$=1 cm$^2$) mounted of a motorized $z$-translation stage (indentation amplitude, $\Delta$z = 1 mm, Figure ~3b-e), while measuring the applied normal force $F$ (Methods Section). With our fluorescence macroscope, we image a pseudo-vesicle containing carboxyfluorescein, as the pseudo-vesicle is cyclically deformed (Figure ~3b-d). Using image analysis, we fit an ellipse to the pseudo-vesicle shape (excluding the oil cap from the analysis, Figure ~3c). The time variation of the long $a$ and short $b$ axis of the ellipse are plotted on Figure ~3f, from which we compute the ellipse's eccentricity $e = \sqrt{1-b^2/a^2}$. The variation of eccentricity, $\Delta e = e(F\ne 0)-e(F=0)$ can be used as a proxy for strain. On Figure ~3g, we plot the compressive stress $\sigma= F/S$ as a function of $e$. A linear relationship is observed, $\sigma = K_e \Delta e$, with an effective compression modulus of the pseudo-vesicle $K_e = 12.1\pm 0.6$ kPa. The value of $K_e$ is well distinct from the gel's compression modulus $K_{gel} \approx 85$ kPa (see inset of Figure  3g). Our method thus offer a handful platform to study lipid bilayer mechanics or mechanotransduction processes.

\begin{figure}[!h]
\centering
\includegraphics[width=0.45\textwidth]{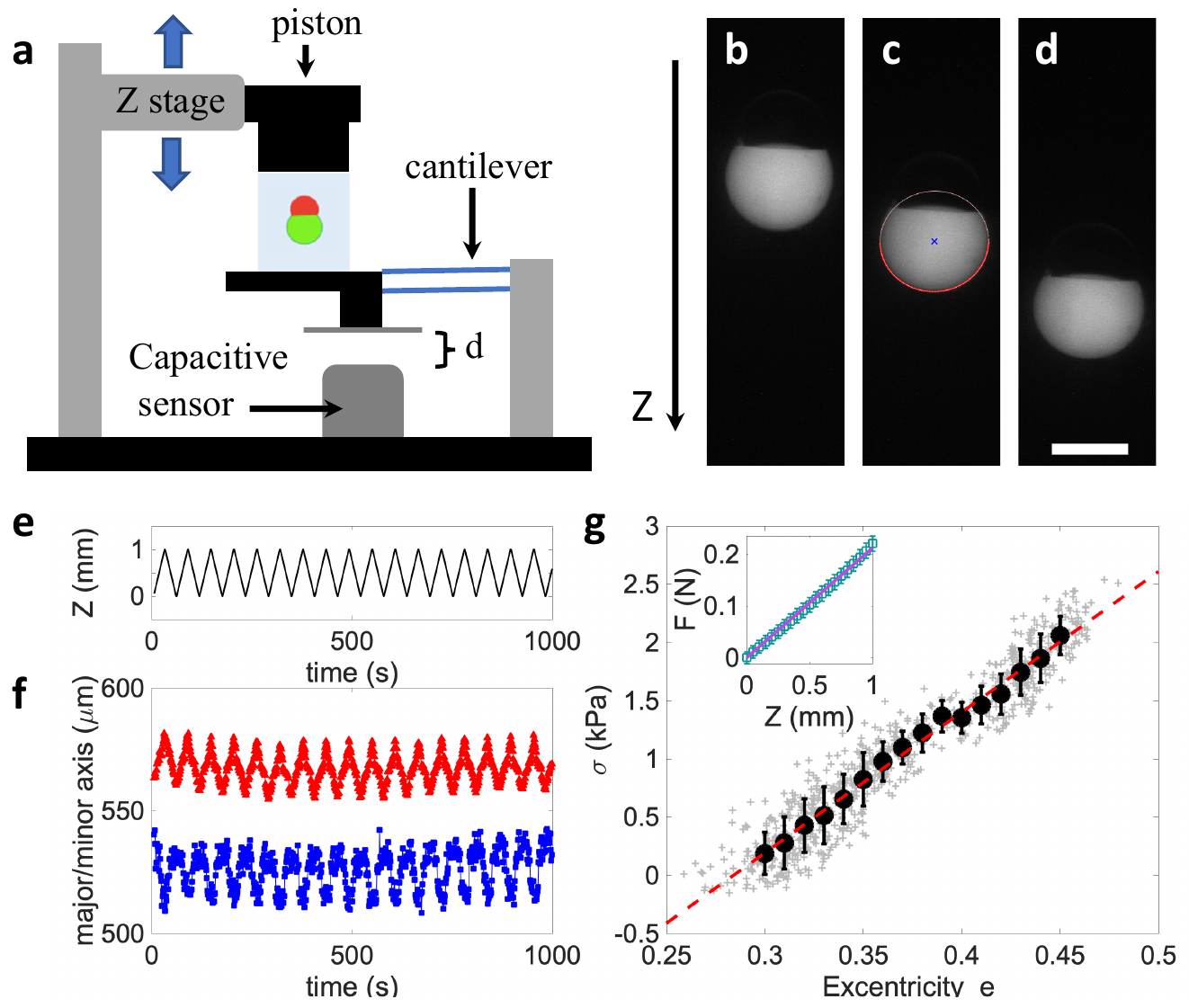}
\caption{ a) Sketch of the mechanical excitation setup. b-d) Macroscope fluorescence images of pseudo-vesicles either relaxed ((b) $Z$=0) or deformed ($Z$=0.5 and 1 mm in (c,d), respectively). In (c) the red line is the fit of the contour of the bottom part of the pseudo-vesicle with an ellipse. Scale bar = 400 $\mu$m. e) Piston position $Z(t)$. f) Major (red) and minor (blue) axes of the fitted ellipse as a function of time, for a 1000 s long cyclic indentation. g) Compressive stress $\sigma$ as a function of the ellipse's eccentricity $e$. Grey crosses correspond to all data points from f), black circles are averages within bins of eccentricity $\delta e$= 0.01. Error bars are SD of the data. The red doted line is a linear fit to the data. Inset: Normal force as a function of the piston indentation. The line is a linear fit $F=\kappa z$, from which the gel compression modulus can be deduced, $K_{gel}=\kappa H/S$, with $H$ the height of the gel.}
\label{fig:fig3}
\end{figure}

\section{Conclusion}
\par Overall our approach constitutes a very simple method to produce and trap pseudo-vesicles within a gel, which is easy to set up and inexpensive. On the one hand, the method only requires small volumes of encapsulated phase ($\sim$ 1 $\mu$L of sample), in contrast with usual microfluidic techniques \cite{li2018microfluidic} which usually require hundreds of microliters of solutions in order to obtain stable flows. On the other hand, it is based on gentle manipulation, thus avoiding any protein denaturation which can be caused by the application of electric fields as in electroformation techniques. It makes it of special interest for the use of valuable and delicate biological samples. Furthermore, because the pseudo-vesicles are trapped in a gel, they are easily localized and do not require post-production handling, in contrast with emulsion template methods. The pseudo-vesicles can also be easily deformed by indenting the gel surface, allowing to finely tune both the stress amplitude and frequency. Finally, the pseudo-vesicles are embedded in a viscoelastic gel which better mimics the mechanical properties of biological tissues.
% Methods Section

\section{Methods Section}
\subsection{Chemicals}
Unless specified, all chemicals were purchased from Sigma Aldrich, Merk inc.
\subsubsection*{Internal/external aqueous phase} 
The internal aqueous phase of the double droplet contains 10 mM Tris buffer (pH=7.5), 400 mM sucrose, and 100 mg/mL Dextran (M$_w$=40.10$^3$ g/mol,). For fluorescent pseudo-vesicles, we added 20 $\mu$M of carboxyfluorescerin to this buffer. The osmolarity of the internal phase, measured with a L{\"o}ser TYP6 osmometer is about 600 mOsm.\\
The external phase is made of a 3\%wt low gelling temperature Agar, dissolved in a buffer containing 10 mM Tris (pH=7.5) and 200 mM potassium chloride. Prior to agar addition we adjust the osmolarity to 380 mosm. Agar is then dissolved in the buffer on a hot plate and the resulting solution is maintained in a liquid form at 85$^\circ$C. 

\subsubsection{Oil phase containing the phospholipids} 

The oil phase is a 50/50 wt/wt mixture of Hexadecan and Silicon oil AR20 in which dried lipids are dissolved. The lipid mixture is the one described in \cite{dupin2019signalling} to mimic the bacterial membrane composition and maximize its stability in droplet interface bilayer geometries. It consists of 1,2-dioleoyl-sn-glycero-3-phosphocholine (DOPC, 81.1 wt\%), 1,2-diphytanoyl-sn-glycero-3-phosphocholine (DPhPC, 10.8 wt\%), 1,2-dioleoyl-sn-glycero-3-phospho-(1'-rac-glycerol) (sodium salt) (DOPG, 5.4 wt\%) and cholesterol (2.7 wt\%). In practice, we prepare a solution of 7.4 mg of this lipid mixture dissolved in chloroform. The lipids are then dried under nitrogen and kept under inert atmosphere in a 2 mL vial at -20$^\circ$C  for several weeks. Immediately before use, we add 2 mL of the oil phase to yield a total lipid concentration of 3.7 mg/mL, and place the vial in an ultrasonic bath for 30 minutes at 30 $^\circ$C. This oil/lipid mixture can then be used for a few days.

For confocal imaging, we labelled separately the oil phase and the phospholipids. The oil phase is labelled with Nile Red with the following procedure: a small amount of solid Nile Red (typically the tip of a spatula) is dissolved in 300 $\mu$L of acetone. 5 mL of silicone oil is then poured onto the acetone and stirred overnight at room temperature in order to transfer the Nile Red dye into the oil phase while evaporating all traces of solvent.

For the labelling of the lipid bilayer, fluorescent lipids were incorporated in the lipid mix following the above-mentioned procedure. We use 1-Myristoyl-2-[12-[(7-nitro-2-1,3-benzoxadiazol-4-yl)amino]dodecanoyl]-sn-Glycero-3-Phosphocholine or NBD-PC lipids that are excited at 480 nm, allowing us to image simultaneously the oil phase labelled with Nile Red and the lipid bilayer labelled with NBD-PC. In that case the final lipid concentrations in the mix yield: 80.3wt\% DOPC, 10.7wt\% DPhPC, 2.7wt\% Cholesterol, 5.4wt\% DOPG, 1wt\% NBD-PC. \\

\subsubsection{SUV preparation}
The above-mentioned lipid mixture (total lipid mass 1.25 mg) is dried under nitrogen in a test tube and left to dry further in a vacuum chamber overnight. The next day,  1 mL of the internal aqueous phase is added to the lipid film and the solution is placed in a probe sonicator for 30 minutes using on/off cycles of 15/5 seconds, respectively. 

\subsubsection{Preparation of $\alpha$HL solutions and integration in SUV}
$\alpha$HL monomers are prepared in the internal aqueous buffer at a concentration of 250 $\mu$g/mL. In order to obtain SUV decorated with $\alpha$HL nanopores, 30 $\mu$L of this $\alpha$HL solution was added to 120 $\mu$L of the SUV solution, yielding a final pore monomer concentration of 50 $\mu$g/mL. The resulting mix is incubated at room temperature for about an hour and used as the inner phase of the pseudo-vesicles.

\subsection{Vesicle formation and mechanical excitation}
The oil/aqueous phase sandwich is made following the steps described in the main text. The warm agar solution is poured into a spectrophotometer cuvette (dimensions 10x10x40 mm$^{3}$). Temperature of the agar is measured with a National Instrument temperature thermocouple (USB-TC01). When the temperature reaches about 38$^\circ$C, the double droplet is formed. The system is left to cool down to room temperature for about 15 minutes, so that the agar gels around the pseudo-vesicle. \\
For mechanical excitation experiments, a thin rectangular sheet of Plexiglas (width 1 mm) is added on one side of the cuvette to cover its wall, prior to liquid agar pouring. As the agar is gelling, this sheet is carefully removed, leaving an empty space between the agar gel and the cuvette wall. Indeed, due to Poisson effect, a lateral expansion of the gel occurs as it is vertically compressed. This empty space is required to allow for this lateral expansion and a proper elastic deformation of the gel.\\
The cuvette is placed and fixed on the device displayed in Figure~3a. The piston has been 3D printed to fit the cuvette size and a thin Plexiglas sheet is glued to the piston face indenting the agar surface. The piston is mounted on a translation stage controlled by a Newport LTA-HL linear actuator ($Z$ resolution of 1 $\mu$m). As the piston indents the gel, it deflects the set of two planar cantilevers attached to the base of the sample holder. A capacitive sensor measures the deflection of the cantilevers. Knowing the stiffness of these cantilever (from previous calibration) we deduce the applied normal force $F$ (the measurement noise on $F$ is 1 mN). \\
The first step of the indentation experiments consists in determining the gel surface position. To this end, the $Z$ position of the piston is lowered by steps of 0.1 mm, while measuring the normal force. As the force reaches 5 mN the ramp is stopped and this $Z$ position is taken as the origin for Z coordinates.\\
Subsequently, we impose a cyclic deformation of the gels, by indenting the gel, from this surface position, by a value $\Delta Z$, by steps of 100 $\mu$m. At each step, the force is measured. The pseudo-vesicle is imaged at each indention step (using a pulse of blue light, duration 500 ms) with a Leica Macroscope and a Pointgrey camera (BFLY-U3-23S6C-C). 

\subsection{Smaller pseudo-vesicle production}
To reach smaller pseudo-vesicle sizes, we use a smaller injection tip. In practice we use a Polyurethan tubing (Phymep) with an outer diameter of 240 $\mu$m and an inner diameter of 130 $\mu$m. The tubing is connected to a 50 $\mu$L syringe (Hamilton), mounted on a home-made micro-injector. The pseudo-vesicles are produced in agar gel with the fluorescent inner buffer and imaged through epifluorescence. We determine their sizes using image analysis and find an average diameter $D= 410 \pm 45 \mu$m (N=6).

% Acknowledgements
\medskip
\textbf{Acknowledgements} \par 
The authors thank Mathieu Letrou and Sophie Cribier for their valuable help with SUV preparation. EW acknowledges financial support from Emergence Sorbonne University, LLP acknowledges support from ANR BOAT and from Emergence(s) Ville de Paris.

\medskip

\bibliography{biblio}

\begin{thebibliography}{26}%
\makeatletter
\providecommand \@ifxundefined [1]{%
 \@ifx{#1\undefined}
}%
\providecommand \@ifnum [1]{%
 \ifnum #1\expandafter \@firstoftwo
 \else \expandafter \@secondoftwo
 \fi
}%
\providecommand \@ifx [1]{%
 \ifx #1\expandafter \@firstoftwo
 \else \expandafter \@secondoftwo
 \fi
}%
\providecommand \natexlab [1]{#1}%
\providecommand \enquote  [1]{``#1''}%
\providecommand \bibnamefont  [1]{#1}%
\providecommand \bibfnamefont [1]{#1}%
\providecommand \citenamefont [1]{#1}%
\providecommand \href@noop [0]{\@secondoftwo}%
\providecommand \href [0]{\begingroup \@sanitize@url \@href}%
\providecommand \@href[1]{\@@startlink{#1}\@@href}%
\providecommand \@@href[1]{\endgroup#1\@@endlink}%
\providecommand \@sanitize@url [0]{\catcode `\\12\catcode `\$12\catcode
  `\&12\catcode `\#12\catcode `\^12\catcode `\_12\catcode `\%12\relax}%
\providecommand \@@startlink[1]{}%
\providecommand \@@endlink[0]{}%
\providecommand \url  [0]{\begingroup\@sanitize@url \@url }%
\providecommand \@url [1]{\endgroup\@href {#1}{\urlprefix }}%
\providecommand \urlprefix  [0]{URL }%
\providecommand \Eprint [0]{\href }%
\providecommand \doibase [0]{http://dx.doi.org/}%
\providecommand \selectlanguage [0]{\@gobble}%
\providecommand \bibinfo  [0]{\@secondoftwo}%
\providecommand \bibfield  [0]{\@secondoftwo}%
\providecommand \translation [1]{[#1]}%
\providecommand \BibitemOpen [0]{}%
\providecommand \bibitemStop [0]{}%
\providecommand \bibitemNoStop [0]{.\EOS\space}%
\providecommand \EOS [0]{\spacefactor3000\relax}%
\providecommand \BibitemShut  [1]{\csname bibitem#1\endcsname}%
\let\auto@bib@innerbib\@empty
%</preamble>
\bibitem [{\citenamefont {Wang}\ \emph {et~al.}(2021)\citenamefont {Wang} \emph
  {et~al.}}]{Wang2021}%
  \BibitemOpen
  \bibfield  {author} {\bibinfo {author} {\bibfnamefont {X.}~\bibnamefont
  {Wang}} \emph {et~al.},\ }\href {\doibase 10.1002} {\bibfield  {journal}
  {\bibinfo  {journal} {Advanced materials}\ }\textbf {\bibinfo {volume} {33}}
  (\bibinfo {year} {2021}),\ 10.1002}\BibitemShut {NoStop}%
\bibitem [{\citenamefont {Jeong}\ \emph {et~al.}(2020)\citenamefont {Jeong},
  \citenamefont {Nguyen}, \citenamefont {Kim}, \citenamefont {Ly},\ and\
  \citenamefont {Shin}}]{jeong2020toward}%
  \BibitemOpen
  \bibfield  {author} {\bibinfo {author} {\bibfnamefont {S.}~\bibnamefont
  {Jeong}}, \bibinfo {author} {\bibfnamefont {H.~T.}\ \bibnamefont {Nguyen}},
  \bibinfo {author} {\bibfnamefont {C.~H.}\ \bibnamefont {Kim}}, \bibinfo
  {author} {\bibfnamefont {M.~N.}\ \bibnamefont {Ly}}, \ and\ \bibinfo {author}
  {\bibfnamefont {K.}~\bibnamefont {Shin}},\ }\href@noop {} {\bibfield
  {journal} {\bibinfo  {journal} {Advanced Functional Materials}\ }\textbf
  {\bibinfo {volume} {30}},\ \bibinfo {pages} {1907182} (\bibinfo {year}
  {2020})}\BibitemShut {NoStop}%
\bibitem [{\citenamefont {Noireaux}\ and\ \citenamefont
  {Libchaber}(2004)}]{noireaux2004vesicle}%
  \BibitemOpen
  \bibfield  {author} {\bibinfo {author} {\bibfnamefont {V.}~\bibnamefont
  {Noireaux}}\ and\ \bibinfo {author} {\bibfnamefont {A.}~\bibnamefont
  {Libchaber}},\ }\href@noop {} {\bibfield  {journal} {\bibinfo  {journal}
  {Proceedings of the National Academy of Sciences}\ }\textbf {\bibinfo
  {volume} {101}},\ \bibinfo {pages} {17669} (\bibinfo {year}
  {2004})}\BibitemShut {NoStop}%
\bibitem [{\citenamefont {Niederholtmeyer}\ \emph {et~al.}(2018)\citenamefont
  {Niederholtmeyer}, \citenamefont {Chaggan},\ and\ \citenamefont
  {Devaraj}}]{niederholtmeyer2018communication}%
  \BibitemOpen
  \bibfield  {author} {\bibinfo {author} {\bibfnamefont {H.}~\bibnamefont
  {Niederholtmeyer}}, \bibinfo {author} {\bibfnamefont {C.}~\bibnamefont
  {Chaggan}}, \ and\ \bibinfo {author} {\bibfnamefont {N.~K.}\ \bibnamefont
  {Devaraj}},\ }\href@noop {} {\bibfield  {journal} {\bibinfo  {journal}
  {Nature communications}\ }\textbf {\bibinfo {volume} {9}},\ \bibinfo {pages}
  {1} (\bibinfo {year} {2018})}\BibitemShut {NoStop}%
\bibitem [{\citenamefont {Pontani}\ \emph {et~al.}(2009)\citenamefont {Pontani}
  \emph {et~al.}}]{pontani2009reconstitution}%
  \BibitemOpen
  \bibfield  {author} {\bibinfo {author} {\bibfnamefont {L.-L.}\ \bibnamefont
  {Pontani}} \emph {et~al.},\ }\href@noop {} {\bibfield  {journal} {\bibinfo
  {journal} {Biophysical journal}\ }\textbf {\bibinfo {volume} {96}},\ \bibinfo
  {pages} {192} (\bibinfo {year} {2009})}\BibitemShut {NoStop}%
\bibitem [{\citenamefont {Gavriljuk}\ \emph {et~al.}(2021)\citenamefont
  {Gavriljuk} \emph {et~al.}}]{Gavriljuk2021}%
  \BibitemOpen
  \bibfield  {author} {\bibinfo {author} {\bibfnamefont {K.}~\bibnamefont
  {Gavriljuk}} \emph {et~al.},\ }\href {\doibase 10.1038/s41467-021-21679-2}
  {\bibfield  {journal} {\bibinfo  {journal} {Nature Com. 2021 12:1}\ }\textbf
  {\bibinfo {volume} {12}},\ \bibinfo {pages} {1} (\bibinfo {year}
  {2021})}\BibitemShut {NoStop}%
\bibitem [{\citenamefont {Chan}\ \emph {et~al.}(2008)\citenamefont {Chan},
  \citenamefont {van Lengerich},\ and\ \citenamefont {Boxer}}]{chan2008lipid}%
  \BibitemOpen
  \bibfield  {author} {\bibinfo {author} {\bibfnamefont {Y.-H.~M.}\
  \bibnamefont {Chan}}, \bibinfo {author} {\bibfnamefont {B.}~\bibnamefont {van
  Lengerich}}, \ and\ \bibinfo {author} {\bibfnamefont {S.~G.}\ \bibnamefont
  {Boxer}},\ }\href@noop {} {\bibfield  {journal} {\bibinfo  {journal}
  {Biointerphases}\ }\textbf {\bibinfo {volume} {3}},\ \bibinfo {pages} {FA17}
  (\bibinfo {year} {2008})}\BibitemShut {NoStop}%
\bibitem [{\citenamefont {Chan}\ \emph {et~al.}(2009)\citenamefont {Chan},
  \citenamefont {van Lengerich},\ and\ \citenamefont
  {Boxer}}]{chan2009effects}%
  \BibitemOpen
  \bibfield  {author} {\bibinfo {author} {\bibfnamefont {Y.-H.~M.}\
  \bibnamefont {Chan}}, \bibinfo {author} {\bibfnamefont {B.}~\bibnamefont {van
  Lengerich}}, \ and\ \bibinfo {author} {\bibfnamefont {S.~G.}\ \bibnamefont
  {Boxer}},\ }\href@noop {} {\bibfield  {journal} {\bibinfo  {journal}
  {Proceedings of the National Academy of Sciences}\ }\textbf {\bibinfo
  {volume} {106}},\ \bibinfo {pages} {979} (\bibinfo {year}
  {2009})}\BibitemShut {NoStop}%
\bibitem [{\citenamefont {Heuvingh}\ \emph {et~al.}(2004)\citenamefont
  {Heuvingh}, \citenamefont {Pincet},\ and\ \citenamefont
  {Cribier}}]{heuvingh2004hemifusion}%
  \BibitemOpen
  \bibfield  {author} {\bibinfo {author} {\bibfnamefont {J.}~\bibnamefont
  {Heuvingh}}, \bibinfo {author} {\bibfnamefont {F.}~\bibnamefont {Pincet}}, \
  and\ \bibinfo {author} {\bibfnamefont {S.}~\bibnamefont {Cribier}},\
  }\href@noop {} {\bibfield  {journal} {\bibinfo  {journal} {The European
  Physical Journal E}\ }\textbf {\bibinfo {volume} {14}},\ \bibinfo {pages}
  {269} (\bibinfo {year} {2004})}\BibitemShut {NoStop}%
\bibitem [{\citenamefont {Garamella}\ \emph {et~al.}(2019)\citenamefont
  {Garamella} \emph {et~al.}}]{Garamella2019}%
  \BibitemOpen
  \bibfield  {author} {\bibinfo {author} {\bibfnamefont {J.}~\bibnamefont
  {Garamella}} \emph {et~al.},\ }\href {\doibase
  10.1021/ACSSYNBIO.9B00204/SUPPL_FILE/SB9B00204_SI_001.PDF} {\bibfield
  {journal} {\bibinfo  {journal} {ACS Synth. Biol.}\ }\textbf {\bibinfo
  {volume} {8}},\ \bibinfo {pages} {1913} (\bibinfo {year} {2019})}\BibitemShut
  {NoStop}%
\bibitem [{\citenamefont {Hindley}\ \emph {et~al.}(2019)\citenamefont {Hindley}
  \emph {et~al.}}]{Hindley2019}%
  \BibitemOpen
  \bibfield  {author} {\bibinfo {author} {\bibfnamefont {J.~W.}\ \bibnamefont
  {Hindley}} \emph {et~al.},\ }\href {\doibase
  10.1073/PNAS.1903500116/-/DCSUPPLEMENTAL} {\bibfield  {journal} {\bibinfo
  {journal} {PNAS}\ }\textbf {\bibinfo {volume} {116}},\ \bibinfo {pages}
  {16711} (\bibinfo {year} {2019})}\BibitemShut {NoStop}%
\bibitem [{\citenamefont {Bangham}\ and\ \citenamefont
  {Horne}(1964)}]{bangham1964negative}%
  \BibitemOpen
  \bibfield  {author} {\bibinfo {author} {\bibfnamefont {A.~D.}\ \bibnamefont
  {Bangham}}\ and\ \bibinfo {author} {\bibfnamefont {R.}~\bibnamefont
  {Horne}},\ }\href@noop {} {\bibfield  {journal} {\bibinfo  {journal} {Journal
  of molecular biology}\ }\textbf {\bibinfo {volume} {8}},\ \bibinfo {pages}
  {660} (\bibinfo {year} {1964})}\BibitemShut {NoStop}%
\bibitem [{\citenamefont {Stein}\ \emph {et~al.}(2017)\citenamefont {Stein},
  \citenamefont {Spindler}, \citenamefont {Bonakdar}, \citenamefont {Wang},\
  and\ \citenamefont {Sandoghdar}}]{stein2017production}%
  \BibitemOpen
  \bibfield  {author} {\bibinfo {author} {\bibfnamefont {H.}~\bibnamefont
  {Stein}}, \bibinfo {author} {\bibfnamefont {S.}~\bibnamefont {Spindler}},
  \bibinfo {author} {\bibfnamefont {N.}~\bibnamefont {Bonakdar}}, \bibinfo
  {author} {\bibfnamefont {C.}~\bibnamefont {Wang}}, \ and\ \bibinfo {author}
  {\bibfnamefont {V.}~\bibnamefont {Sandoghdar}},\ }\href@noop {} {\bibfield
  {journal} {\bibinfo  {journal} {Frontiers in physiology}\ }\textbf {\bibinfo
  {volume} {8}},\ \bibinfo {pages} {63} (\bibinfo {year} {2017})}\BibitemShut
  {NoStop}%
\bibitem [{\citenamefont {Angelova}\ and\ \citenamefont
  {Dimitrov}(1986)}]{angelova1986liposome}%
  \BibitemOpen
  \bibfield  {author} {\bibinfo {author} {\bibfnamefont {M.~I.}\ \bibnamefont
  {Angelova}}\ and\ \bibinfo {author} {\bibfnamefont {D.~S.}\ \bibnamefont
  {Dimitrov}},\ }\href@noop {} {\bibfield  {journal} {\bibinfo  {journal}
  {Faraday discussions of the Chemical Society}\ }\textbf {\bibinfo {volume}
  {81}},\ \bibinfo {pages} {303} (\bibinfo {year} {1986})}\BibitemShut
  {NoStop}%
\bibitem [{\citenamefont {Pautot}\ \emph {et~al.}(2003)\citenamefont {Pautot},
  \citenamefont {Frisken},\ and\ \citenamefont {Weitz}}]{pautot2003production}%
  \BibitemOpen
  \bibfield  {author} {\bibinfo {author} {\bibfnamefont {S.}~\bibnamefont
  {Pautot}}, \bibinfo {author} {\bibfnamefont {B.~J.}\ \bibnamefont {Frisken}},
  \ and\ \bibinfo {author} {\bibfnamefont {D.}~\bibnamefont {Weitz}},\
  }\href@noop {} {\bibfield  {journal} {\bibinfo  {journal} {Langmuir}\
  }\textbf {\bibinfo {volume} {19}},\ \bibinfo {pages} {2870} (\bibinfo {year}
  {2003})}\BibitemShut {NoStop}%
\bibitem [{\citenamefont {Li}\ \emph {et~al.}(2018)\citenamefont {Li},
  \citenamefont {Zhang}, \citenamefont {Ge}, \citenamefont {Xu}, \citenamefont
  {Zhang}, \citenamefont {Qu}, \citenamefont {Choi}, \citenamefont {Xu},
  \citenamefont {Zhang}, \citenamefont {Lee} \emph
  {et~al.}}]{li2018microfluidic}%
  \BibitemOpen
  \bibfield  {author} {\bibinfo {author} {\bibfnamefont {W.}~\bibnamefont
  {Li}}, \bibinfo {author} {\bibfnamefont {L.}~\bibnamefont {Zhang}}, \bibinfo
  {author} {\bibfnamefont {X.}~\bibnamefont {Ge}}, \bibinfo {author}
  {\bibfnamefont {B.}~\bibnamefont {Xu}}, \bibinfo {author} {\bibfnamefont
  {W.}~\bibnamefont {Zhang}}, \bibinfo {author} {\bibfnamefont
  {L.}~\bibnamefont {Qu}}, \bibinfo {author} {\bibfnamefont {C.-H.}\
  \bibnamefont {Choi}}, \bibinfo {author} {\bibfnamefont {J.}~\bibnamefont
  {Xu}}, \bibinfo {author} {\bibfnamefont {A.}~\bibnamefont {Zhang}}, \bibinfo
  {author} {\bibfnamefont {H.}~\bibnamefont {Lee}},  \emph {et~al.},\
  }\href@noop {} {\bibfield  {journal} {\bibinfo  {journal} {Chemical Society
  Reviews}\ }\textbf {\bibinfo {volume} {47}},\ \bibinfo {pages} {5646}
  (\bibinfo {year} {2018})}\BibitemShut {NoStop}%
\bibitem [{\citenamefont {Evans}\ and\ \citenamefont
  {Needham}(1987)}]{Evans1987}%
  \BibitemOpen
  \bibfield  {author} {\bibinfo {author} {\bibfnamefont {E.}~\bibnamefont
  {Evans}}\ and\ \bibinfo {author} {\bibfnamefont {D.}~\bibnamefont
  {Needham}},\ }\href {\doibase 10.1021/j100300a003} {\bibfield  {journal}
  {\bibinfo  {journal} {J. Phys. Chem.}\ }\textbf {\bibinfo {volume} {91}},\
  \bibinfo {pages} {4219} (\bibinfo {year} {1987})}\BibitemShut {NoStop}%
\bibitem [{\citenamefont {Kulin}\ \emph {et~al.}(2003)\citenamefont {Kulin}
  \emph {et~al.}}]{Kulin2003}%
  \BibitemOpen
  \bibfield  {author} {\bibinfo {author} {\bibfnamefont {S.}~\bibnamefont
  {Kulin}} \emph {et~al.},\ }\href {\doibase 10.1021/la0344433} {\bibfield
  {journal} {\bibinfo  {journal} {Langmuir}\ }\textbf {\bibinfo {volume}
  {19}},\ \bibinfo {pages} {8206} (\bibinfo {year} {2003})}\BibitemShut
  {NoStop}%
\bibitem [{\citenamefont {Garten}\ \emph {et~al.}(2017)\citenamefont {Garten}
  \emph {et~al.}}]{Garten2017}%
  \BibitemOpen
  \bibfield  {author} {\bibinfo {author} {\bibfnamefont {M.}~\bibnamefont
  {Garten}} \emph {et~al.},\ }\href {\doibase 10.1073/pnas.1609142114}
  {\bibfield  {journal} {\bibinfo  {journal} {PNAS}\ }\textbf {\bibinfo
  {volume} {114}},\ \bibinfo {pages} {328} (\bibinfo {year}
  {2017})}\BibitemShut {NoStop}%
\bibitem [{\citenamefont {Robinson}(2019)}]{robinson2019microfluidic}%
  \BibitemOpen
  \bibfield  {author} {\bibinfo {author} {\bibfnamefont {T.}~\bibnamefont
  {Robinson}},\ }\href@noop {} {\bibfield  {journal} {\bibinfo  {journal}
  {Advanced Biosystems}\ }\textbf {\bibinfo {volume} {3}},\ \bibinfo {pages}
  {1800318} (\bibinfo {year} {2019})}\BibitemShut {NoStop}%
\bibitem [{\citenamefont {Deschamps}\ \emph {et~al.}(2009)\citenamefont
  {Deschamps} \emph {et~al.}}]{deschamps2009dynamics}%
  \BibitemOpen
  \bibfield  {author} {\bibinfo {author} {\bibfnamefont {J.}~\bibnamefont
  {Deschamps}} \emph {et~al.},\ }\href@noop {} {\bibfield  {journal} {\bibinfo
  {journal} {PNAS}\ }\textbf {\bibinfo {volume} {106}},\ \bibinfo {pages}
  {11444} (\bibinfo {year} {2009})}\BibitemShut {NoStop}%
\bibitem [{\citenamefont {Sch{\"a}fer}\ \emph {et~al.}(2015)\citenamefont
  {Sch{\"a}fer} \emph {et~al.}}]{Schafer2015}%
  \BibitemOpen
  \bibfield  {author} {\bibinfo {author} {\bibfnamefont {E.}~\bibnamefont
  {Sch{\"a}fer}} \emph {et~al.},\ }\href {\doibase 10.1039/C5SM00191A}
  {\bibfield  {journal} {\bibinfo  {journal} {Soft Matter}\ }\textbf {\bibinfo
  {volume} {11}},\ \bibinfo {pages} {4487} (\bibinfo {year}
  {2015})}\BibitemShut {NoStop}%
\bibitem [{\citenamefont {Lin}\ \emph {et~al.}(2019)\citenamefont {Lin} \emph
  {et~al.}}]{Lin2019}%
  \BibitemOpen
  \bibfield  {author} {\bibinfo {author} {\bibfnamefont {Y.~C.}\ \bibnamefont
  {Lin}} \emph {et~al.},\ }\href {\doibase 10.1038/S41586-019-1499-2}
  {\bibfield  {journal} {\bibinfo  {journal} {Nature}\ }\textbf {\bibinfo
  {volume} {573}},\ \bibinfo {pages} {230} (\bibinfo {year}
  {2019})}\BibitemShut {NoStop}%
\bibitem [{\citenamefont {Gouaux}\ \emph {et~al.}(1994)\citenamefont {Gouaux}
  \emph {et~al.}}]{gouaux1994subunit}%
  \BibitemOpen
  \bibfield  {author} {\bibinfo {author} {\bibfnamefont {J.~E.}\ \bibnamefont
  {Gouaux}} \emph {et~al.},\ }\href@noop {} {\bibfield  {journal} {\bibinfo
  {journal} {PNAS}\ }\textbf {\bibinfo {volume} {91}},\ \bibinfo {pages}
  {12828} (\bibinfo {year} {1994})}\BibitemShut {NoStop}%
\bibitem [{\citenamefont {Valet}\ \emph {et~al.}(2019)\citenamefont {Valet}
  \emph {et~al.}}]{valet2019diffusion}%
  \BibitemOpen
  \bibfield  {author} {\bibinfo {author} {\bibfnamefont {M.}~\bibnamefont
  {Valet}} \emph {et~al.},\ }\href@noop {} {\bibfield  {journal} {\bibinfo
  {journal} {PRL}\ }\textbf {\bibinfo {volume} {123}},\ \bibinfo {pages}
  {088101} (\bibinfo {year} {2019})}\BibitemShut {NoStop}%
\bibitem [{\citenamefont {Dupin}\ and\ \citenamefont
  {Simmel}(2019)}]{dupin2019signalling}%
  \BibitemOpen
  \bibfield  {author} {\bibinfo {author} {\bibfnamefont {A.}~\bibnamefont
  {Dupin}}\ and\ \bibinfo {author} {\bibfnamefont {F.~C.}\ \bibnamefont
  {Simmel}},\ }\href@noop {} {\bibfield  {journal} {\bibinfo  {journal} {Nature
  chem.}\ }\textbf {\bibinfo {volume} {11}},\ \bibinfo {pages} {32} (\bibinfo
  {year} {2019})}\BibitemShut {NoStop}%
\end{thebibliography}%

\end{document}